\documentclass[printer]{tMOP2e}
\usepackage{amsmath,amsthm,amsfonts,amssymb,epsfig}

\begin{document}

\doi{}

\markboth{M. D. Grace \emph{et al.}}{Fidelity of optimally controlled
quantum gates}

\title{Fidelity of optimally controlled quantum gates with randomly
  coupled multiparticle environments}

\author{MATTHEW~D.~GRACE$^{\dagger}$, CONSTANTIN~BRIF$^{\dagger}$,
  HERSCHEL~RABITZ$^{\dagger}$, DANIEL~A.~LIDAR$^{\ddagger}$
  IAN~A.~WALMSLEY$^{\S}$, and ROBERT~L.~KOSUT$^{\P}$ \\
  \vspace{6pt} $^{\dagger}$ Department of Chemistry, Princeton
  University, Princeton, New Jersey 08544 \\ 
  \vspace{6pt} $^{\ddagger}$ Departments of Chemistry, Electrical
  Engineering, and Physics, \\ University of Southern California, Los
  Angeles, CA 90089 \\
  \vspace{6pt} $^{\S}$ Department of Physics, University of
  Oxford, Oxford OX1 3PU, UK \\
  \vspace{6pt} $^{\P}$ SC Solutions, Inc., 1261 Oakmead Parkway,
  Sunnyvale, CA 94085}

\maketitle


\begin{abstract}
This work studies the feasibility of optimal control of high-fidelity
quantum gates in a model of interacting two-level particles. One
particle (the qubit) serves as the quantum information processor, whose
evolution is controlled by a time-dependent external field. The other
particles are not directly controlled and serve as an effective
environment, coupling to which is the source of decoherence. The control
objective is to generate target one-qubit gates in the presence of
strong environmentally-induced decoherence and under physically
motivated restrictions on the control field. It is found that
interactions among the environmental particles have a negligible effect
on the gate fidelity and require no additional adjustment of the control
field. Another interesting result is that optimally controlled quantum
gates are remarkably robust to random variations in qubit-environment
and inter-environment coupling strengths. These findings demonstrate the
utility of optimal control for management of quantum-information systems
in a very precise and specific manner, especially when the dynamics
complexity is exacerbated by inherently uncertain environmental coupling.
\end{abstract}


\section{Introduction}

The methods of optimal control are very useful for effectively managing
various quantum systems \cite{RdVRMK00, WR03} and are particularly
important in situations requiring precise quantum operations, as is the
case for quantum computation (QC) \cite{NiCh00}. One of the most
difficult problems of QC is that unavoidable coupling of the quantum
information processor (QIP) to the environment results in a loss of
coherence. In recent years, significant attention was devoted to various
methods of dynamic suppression of environmentally-induced decoherence in
open quantum systems, including applications of pre-designed external
fields \cite{VKL99a, VKL99b, VKL00, KK04, Facchi05} and optimal control
techniques \cite{Brif01, ZR03, STN04, GK05a, JP05, JP06, WP06}. In a
separate line of research, several works \cite{SKH99, PK02, PK03,
Khaneja05, Grace06, Sp07} considered the generation of optimally
controlled unitary quantum gates in ideal situations where coupling to
the environment can be neglected during the gate operation.

The optimal control of quantum gates in the presence of decoherence
still remains to be fully explored. Two recent works \cite{GK05b,
Schulte06} discussed specific techniques, involving optimizations over
sets of controls operating in pre-designed ``weak-decoherence''
subspaces. We recently proposed \cite{Grace07} a different approach in
which the full power of optimal control theory is used to generate the
target gate transformation with the highest possible fidelity while
simultaneously suppressing decoherence induced by coupling to a
multiparticle environment. This method does not rely on any special
pre-design of system parameters to weaken decoherence (e.g., using
tunable inter-qubit couplings as in \cite{GK05b} or auxiliary qubits as
in \cite{Schulte06}); the only control used in our approach is a
time-dependent external field. A similar application of optimal control
was also recently considered in \cite{Reben06} for another model of a
decohering environment. Optimization techniques were also applied
recently to quantum error correction (QEC) \cite{RW05, KL06}. In
contrast to QEC, our approach does not require ancilla qubits and is not
limited to the weak decoherence regime. The optimal control of quantum
gates can potentially be used in conjunction with QEC to achieve fault
tolerance with an improved threshold.

In the previous work \cite{Grace07}, we showed that optimal control
fields, found by employing a combination of genetic and gradient
algorithms, are able to produce high-fidelity quantum gates in the
presence of strong decoherence. Optimal solutions revealed interesting
control mechanisms that utilize dynamic Stark shifts to weaken coupling
to the environment and control-induced revivals to restore coherence. In
the present work, we extend the analysis of optimally controlled quantum
gates to situations where (i) the environmental particles interact with
each other and (ii) couplings between the QIP and environment and within
the environment itself have randomly varied strengths. Taking into
account these additional environmental effects makes our model more
closely related to realistic quantum information systems (in particular,
spin-based solid-state realizations of QC \cite{Loss, Kane, Vrijen,
Petta, Lyon}). An interesting finding is that the effect of
inter-environment interactions on the fidelity of optimally controlled
quantum gates is negligible (potentially allowing one to neglect certain
interaction terms in the Hamiltonian). We also demonstrate that the
optimal control generates quantum gates which are inherently robust to
random variations in qubit-environment and inter-environment coupling
strengths.


\section{Model system} 

We use a model of interacting two-level particles (e.g., spin-half
particles or two-level atoms), which are divided into the QIP, composed
of one qubit, and an $n$-particle environment. The qubit is directly
coupled to a time-dependent external control field, while the
environment is not directly controlled and is managed only through its
interaction with the qubit. The evolution of the composite system of the
qubit and environment is treated in an exact quantum-mechanical manner,
without either approximating the dynamics by a master equation or using
a perturbative analysis based on the weak coupling assumption. The
Hamiltonian for the controlled system, $H = H_0 + H_C +
H_{\mathrm{int}}$, has the form $\left( \hbar = 1 \right)$
\begin{equation}
H = \sum_{i = 0}^n \omega_i S_{iz} - \mu C(t)
S_{0x} - \sum_{i < j} \gamma_{ij} \mathbf{S}_i \cdot \mathbf{S}_j.
\label{Ham}
\end{equation}
Here, $i = 0$ labels the qubit and $i = 1,\ldots,n$ label the
environmental particles, $\mathbf{S}_i = \left( S_{ix}, S_{iy}, S_{iz}
\right)$ is the spin operator for the $i$th particle ($\mathbf{S}_i =
\frac{1}{2} \bm{\sigma}_i$, in terms of the Pauli matrices), $H_0$ is
the sum over the free Hamiltonians $\omega_i S_{iz}$ for all $n+1$
particles ($\omega_i$ is the transition angular frequency for the $i$th
particle), $H_C$ specifies the coupling between the qubit and the
time-dependent control field $C(t)$ ($\mu$ is the dipole moment), and
$H_{\mathrm{int}}$ represents the Heisenberg exchange interaction
between the particles ($\gamma_{ij}$ is the coupling parameter for the
$i$th and $j$th particles). This model is particularly relevant to
spin-based solid-state realizations of quantum gates (see, e.g.,
\cite{Loss, Kane, Vrijen, Petta, Lyon}), in which unwanted
interactions exist due to impurities in semiconductor structures or
usage of relatively dense lattices of spin-like qubits (e.g., electron
spins in an array of quantum dots or electrons on liquid helium).

In this work, we optimize one-qubit gates coupled to $n$-particle
environments ($n = 1, 2, 4, 6$). For $n = 2$, the system can be modeled
as a two-dimensional triangular lattice with the qubit $q_0$ coupled to
two environmental particles $e_1$ and $e_2$:
\begin{equation}
\begin{array}{ccc}
& q_0 & \\ 
\stackrel{\gamma_{01}}{} \swarrow \! \! \! \! \! \! \nearrow & &
\nwarrow \! \! \! \! \! \! \searrow \stackrel{\gamma_{02}}{} \\ 
e_1 \ \ & \stackrel{\gamma_{12}}{\longleftrightarrow} & \ \ e_2
\end{array}
\label{triangle}
\end{equation}
For $n = 4$, the system can be modeled as a two-dimensional lattice with
the qubit $q_0$ at the center, coupled to four environmental particles
$\{ e_1 , \ldots , e_4 \}$:
\begin{equation}
\begin{array}{ccccc}
 & & e_3 & & \\ 
 & \swarrow \! \! \! \! \! \! \nearrow & \updownarrow & 
   \nwarrow \! \! \! \! \! \! \searrow & \\ 
e_1 & \longleftrightarrow & q_0 & \longleftrightarrow & e_2 \\
 & \nwarrow \! \! \! \! \! \! \searrow & \updownarrow & 
   \swarrow \! \! \! \! \! \! \nearrow & \\ 
 & & e_4 & &
\end{array}
\label{square}
\end{equation}
Similarly, for $n = 6$, the system can be modeled as a three-dimensional
lattice with the qubit at the center, coupled to six environmental
particles. Since the evolution of the composite system is numerically
exact and our optimization procedure (described in
Section~\ref{optimize} and \cite{Grace07}) is iterative, we limit the
number of environmental particles to $n \leq 6$ for computationally
tractable simulations.

We will first consider the case of well-specified coupling parameters
given by
\begin{equation}
\gamma_{ij} = \left\{ \begin{array}{ll} \gamma, 
& \textrm{for} \ i = 0 \ \textrm{and} \ j = 1, \ldots, n, \\ 
\gamma', & \textrm{for} \ i = 1, \ldots, n-1, \ \textrm{and} \ j > i,
\end{array}\right.
\label{gamma}
\end{equation}
which implies that the qubit interacts with each environmental particle
with the coupling parameter $\gamma$ and environmental particles
interact with each other with the coupling parameter $\gamma'$ (compare
to \cite{Grace07} where $\gamma' = 0$). Then we will consider a more
general situation in which every coupling parameter $\gamma_{ij}$ takes
a random value from a normal distribution.


\section{Distance measure for evolution operators}

Let $U(t) \in \mathrm{U}(2^{n+1})$ be the unitary time-evolution
operator of the composite system and $G \in \mathrm{U}(2)$ be the
unitary target transformation for the quantum gate. The evolution is
governed by the Schr\"{o}dinger equation, $\dot{U}(t) = - i H(t) U(t)$,
with the initial condition $U(0) = I$. The gate fidelity depends on the
distance between the actual evolution $U \equiv U(t_{\mathrm{f}})$ at
the final time $t_{\mathrm{f}}$ and the target transformation $G$. In
order to achieve a perfect gate, it suffices for the time-evolution
operator at $t = t_{\mathrm{f}}$ to be in a tensor-product form
$U_{\mathrm{opt}} = G \otimes \Phi$, where $\Phi \in \mathrm{U}(2^{n})$
is an arbitrary unitary transformation acting on the
environment. Therefore, the following objective functional is proposed
as the measure of the distance between $U$ and $G$ \cite{KGBR06}: $J =
\lambda_n \underset{\Phi}{\min} \| U - G \otimes \Phi \|$ subject to
$\Phi \in \mathrm{U}(2^{n})$ (where $\| \cdot \|$ is a matrix norm on
the space $M_{2^{n+1}} \left( \mathbb{C} \right)$ of $2^{n+1} \times
2^{n+1}$ complex matrices and $\lambda_n$ is a normalization
factor). Using the Frobenius norm, defined as $\| A \|_{\mathrm{Fr}} =
\left[ \mathrm{trace} \left( A^{\dag} A \right) \right]^{1/2}$, and
$\lambda_n = 2^{-(n+2)/2}$, the distance measure becomes \cite{KGBR06}
\begin{gather}
\label{frobdist}
J = \left[ 1 -  2 \lambda_n^2 \mathrm{trace} \left( \sqrt{Q^{\dag}Q}
  \right) \right]^{1/2}, \\ 
Q_{\nu \nu'} = \sum_{r, r' = 1}^{2} G_{r r'}^{*} U_{r r' \nu \nu'},
\end{gather}
where $Q \in M_{2^n}\left( \mathbb{C} \right)$ and $Q_{\nu \nu'}$, $G_{r
r'}$, and $U_{r r' \nu \nu'}$ are elements of the matrix representations
of $Q$, $G$, and $U$, respectively. Since $0 \leq J \leq 1$, it is
convenient to define the gate fidelity as $F = 1 - J$. An important
property of this distance measure is its independence of the initial
state. In contrast to some other distance measures \cite{NiCh00}, $J$ is
evaluated directly from the evolution operator $U$, with no need to
specify the initial state of the system. This property of $J$ reflects
our objective of generating a specified target transformation for
whatever initial state, pure or mixed, direct-product or entangled.


\section{Measure of decoherence and system parameters}

A useful measure of decoherence is the von Neumann entropy:
$S_{\mathrm{vN}}(t) = - \mathrm{trace} \left\{ \rho_{\mathrm{q}}(t) \ln
\left[ \rho_{\mathrm{q}}(t) \right] \right\}$, where
$\rho_{\mathrm{q}}(t)$ is the reduced density matrix for the qubit,
$\rho_{\mathrm{q}}(t) = \mathrm{trace}_{\mathrm{env}} \left( \rho(t)
\right)$. For a pure state, $S_{\mathrm{vN}} = 0$, while for a maximally
mixed state of a $k$-level system, $S_{\mathrm{vN}} = \ln(k)$. The
initial state used for the entropy calculations is $|\Psi(0) \rangle =
|-\rangle_0 \otimes \bigotimes_{i = 1}^{n} |+\rangle_i$, where $S_{iz}
|\pm\rangle_i = \pm \frac{1}{2} |\pm\rangle_i$. The distance measure $J$
of Eq.~(\ref{frobdist}) is independent of the initial state and
consequently so are the optimal control fields found for the target
gates and the corresponding fidelities. Therefore, the specific choice
of the initial state for the entropy calculations places no limitations
whatsoever on the generality of the optimal control results.

For the optimal control simulations below, the system parameters are
chosen to ensure complex dynamics and strong decoherence: values of
$\gamma/\omega$ are up to 0.02, which is significant for QC
applications, and frequencies $\omega_i$ are close (but not equal) to
enhance the interaction. We define the unit of time and introduce a
natural system of units by choosing the qubit frequency $\omega_0 = 1$
for all simulations (implying that one period of free evolution is $2
\upi$). The frequencies of the environmental particles are:  $\omega_1
\approx 0.99841$, $\omega_2 \approx 1.00159$, $\omega_3 \approx
0.96007$, $\omega_4 \approx 1.04159$, $\omega_5 \approx 0.87597$,
$\omega_6 \approx 1.14159$ (see \cite{Grace07} for details). Imposing
upper limits on the gate duration $( t_{\mathrm{f}} \leq 60 )$ and
coupling parameter $( \gamma \leq 0.02 )$ places the dynamics of the
uncontrolled system in the regime where decoherence increases
monotonically with time (before the entropy reaches its maximum value of
$S_{\mathrm{vN}} \approx \ln 2$). This dynamical regime approximates
some of the effects that the QIP would experience from a larger
environment, in particular, preventing restoration of coherence to the
qubit by uncontrolled revivals. Thus, any increase in coherence may be
attributed exclusively to the action of the control field.


\section{Optimization procedure}
\label{optimize}
 
Combined genetic and gradient algorithms are employed to minimize the
distance measure $J$ of Eq.~(\ref{frobdist}) (or, equivalently, to
maximize the fidelity $F$) with respect to the control field $C(t)$. The
target quantum logical transformation is the one-qubit Hadamard gate (an
element of a universal set of logical operations for QC \cite{NiCh00}).

When a genetic algorithm is used, the gate fidelity $F$ is maximized
with respect to a parameterized control field $C(t) = f(t)\sum_{\ell}
A_{\ell} \cos \left( \tilde{\omega}_{\ell} t + \theta_{\ell}
\right)$. Here, $f(t)$ is an envelope function incorporating the field's
spectral width and $A_{\ell}$, $\tilde{\omega}_{\ell}$, and
$\theta_{\ell}$ are the amplitude, central frequency, and relative phase
of the $\ell$th component of the field, respectively. A combination of
these parameters (``genes'') represents an ``individual'' (whose
``fitness'' is the gate fidelity), and a collection of individuals
constitutes a ``population'' (we use population sizes of $\sim 250$).

Removing the constraints on the control field imposed by the
parameterized form above provides the potential for more effective
control of the system. In this case the optimal control field is found
by minimizing the following functional \cite{PK02, PK03}:
\begin{equation}
K = J + \mathrm{Re} \int_0^{t_{\mathrm{f}}} 
\mathrm{trace} \left\{ [\dot{U}(t) + i H(t) U(t)] B(t) \right\} dt 
+ \frac{\alpha}{2} \int_0^{t_{\mathrm{f}}} \left| C(t) \right|^2 dt.
\end{equation}
Upon minimization of $K$, the first integral term constrains $U(t)$ to
obey the Schr\"{o}dinger equation ($B(t)$ is an operator Lagrange
multiplier) and the second integral term penalizes the field fluence
$\mathcal{E} = \int_0^{t_{\mathrm{f}}} \left| C(t) \right|^2 dt$ with a
weight $\alpha > 0$. Applying the calculus of variations to $K$ with
respect to $B(t)$ and $U(t)$ yields the Schr\"{o}dinger equation for
$U(t)$ and the time-reversed Schr\"{o}dinger equation for $B(t)$:
$\dot{B}(t) = iB(t)H(t)$, with an appropriate final time condition. The
optimal field is found iteratively, using a gradient algorithm (see
\cite{Grace07} for optimization details).

Despite the lack of direct coupling of the control field to the
environment, it can be shown that the composite system described by
Eq.~(\ref{Ham}) is completely controllable (up to a global phase), as
defined in \cite{RSDRP95}. However, the restrictions on the gate
duration and on the shape of the control field limit the achievable
fidelity.


\section{Results: Optimally controlled one-qubit gate with
  multiparticle environments}

\begin{figure}[!t]
\epsfxsize=0.52\textwidth \centerline{\epsfbox{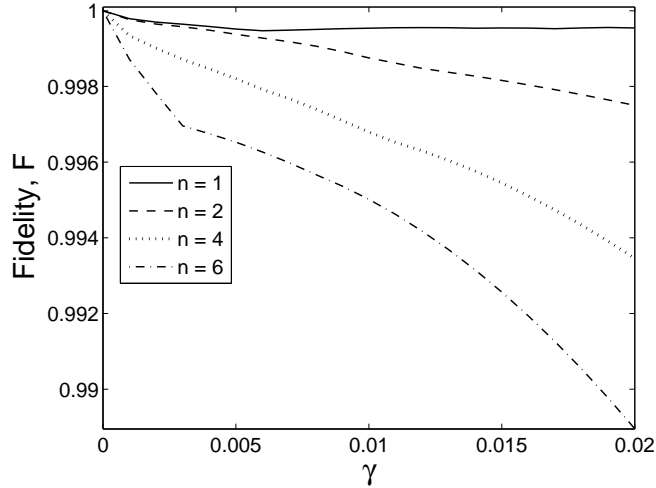}}
\caption{The gate fidelity $F$ versus the qubit-environment coupling
parameter $\gamma$, for the Hadamard gates optimally controlled in the
presence of $n$-particle environments ($n = 1, 2, 4, 6$). Values of
$\gamma$ range from 0 to 0.02 in increments of 0.001.}
\label{fig:fid}
\end{figure}

Fidelities of one-qubit Hadamard gates optimized in the presence of
$n$-particle environments ($n = 1, 2, 4, 6$) are presented in
figure~\ref{fig:fid} for $\gamma' = 0$ and various values of the
qubit-environment coupling parameter $\gamma$. For a one-particle
environment, the control fields optimized for the actual values of
$\gamma$ result in fidelities at least above 0.9994. In particular, we
obtain $F > 1 - 10^{-6}$ for $\gamma = 0$ (a closed system) and $F
\approx 0.9995$ for $\gamma = 0.02$ (the strongest coupling
considered). However, it becomes more difficult to counteract
decoherence as the number of the environmental particles increases; as
seen in figure~\ref{fig:fid}, for larger values of $n$ the gate fidelity
decreases more rapidly as $\gamma$ increases. We also find that applying
a control field optimized for a one-particle environment ($n = 1$) to
$n \geq 2$ results in a significant fidelity loss (up to ten percent
of the original value). This demonstrates the dependence of the
optimal control on the size of the environment.

Optimal control field parameters, gate fidelity, and final-time entropy
for the Hadamard gate coupled to $n$-particle environments ($n = 1, 2,
4, 6$, $\gamma = 0.02$, and $\gamma' = 0$) are reported in
table~\ref{tab:results}. The fields are intense, with maximum amplitudes
ranging from approximately 2.0 (for $n = 1$) to 3.9 (for $n = 2$). The
exact time structure of the optimal field is not intuitive and is
delicately adjusted to the particular control application. For example,
control fields optimized for $\gamma = 0.02$ are not only more intense
than those optimized for $\gamma = 0$, they also have very different
structures. We also find that high-fidelity optimal solutions for
$\gamma = 0$ are obtained for control pulse durations $t_{\mathrm{f}}
\leq 12.0$. In comparison, for $\gamma = 0.02$, high-fidelity optimal
solutions are obtained at longer pulse durations (cf. results reported
in table~\ref{tab:results}, e.g., $t_{\mathrm{f}} = 15.4$ for $n = 2$
and $t_{\mathrm{f}} = 25.0$ for $n = 1$ and $n = 4$). When the
qubit-environment interaction is on, the control field is required to
generate the target gate transformation and at the same time counteract
decoherence. As described below, the control accomplishes the latter
goal by restoring coherence to the QIP and therefore longer pulse
durations are needed in the presence of environmental coupling. As
mentioned above, there are also significant differences in the control
fields optimized for different numbers of environmental particles.

\begin{table}
\tbl{Optimal control field parameters (the maximum field amplitude
$A_{\mathrm{max}}$, control duration $t_{\mathrm{f}}$, and field fluence
$\mathcal{E}$), gate fidelity $F$, and final-time entropy
$S_{\mathrm{vN}}(t_{\mathrm{f}})$ for the Hadamard one-qubit gates
coupled to $n$-particle environments ($n = 1, 2, 4, 6$ and $\gamma =
0.02$). The initial state for the entropy computation is $|\Psi(0)
\rangle$.}
{\begin{tabular}{lcccc}
\toprule
$n$ & 1 & 2 & 4 & 6 \\
\colrule
$A_{\mathrm{max}}$ & 2.0 & 3.9 & 3.8 & 3.1 \\
$t_{\mathrm{f}}$ & 25.0 & 15.4 & 25.0 & 30.0 \\
$\mathcal{E}$ & 20.0 & 49.0 & 55.5 & 54.5 \\
\colrule
$F$ & 0.9995 & 0.9975 & 0.9935 & 0.9890 \\
$S_{\mathrm{vN}}(t_{\mathrm{f}})$ & \ \ $8.6 \times 10^{-8}$ \ & \ 
$4.4 \times 10^{-5}$ \ & \ $4.7 \times 10^{-4}$ \ & \ 
$2.4 \times 10^{-3}$ \ \\
\botrule
\end{tabular}}
\label{tab:results}
\end{table}

We also would like to explore how interactions among the environmental
particles affect optimally controlled gate operations. Interestingly, we
find that for a given $n$-particle environment ($n \geq 2$ and $\gamma =
0.02$), some optimal control solutions obtained for $\gamma' \neq 0$
(e.g., $\gamma' = \frac{7}{8}\gamma = 0.0175$) are essentially identical
to the solutions found for $\gamma' = 0$. This means that no additional
adjustment of the control field is necessary to account for the effect
of inter-environment couplings. Moreover, applying these optimal control
fields to the systems with $\gamma' \neq 0$ yields approximately the
same gate fidelity, as seen for $\gamma' = 0$ (the actual decreases in
the fidelity observed for $\gamma' = \frac{1}{2}\gamma$ and $\gamma' =
\frac{7}{8}\gamma$ are of the order of $10^{-5}$ for $n = 2$ and
$10^{-4}$ for $n = 4$).

\begin{figure}[!t]
\epsfxsize=0.62\textwidth \centerline{\epsfbox{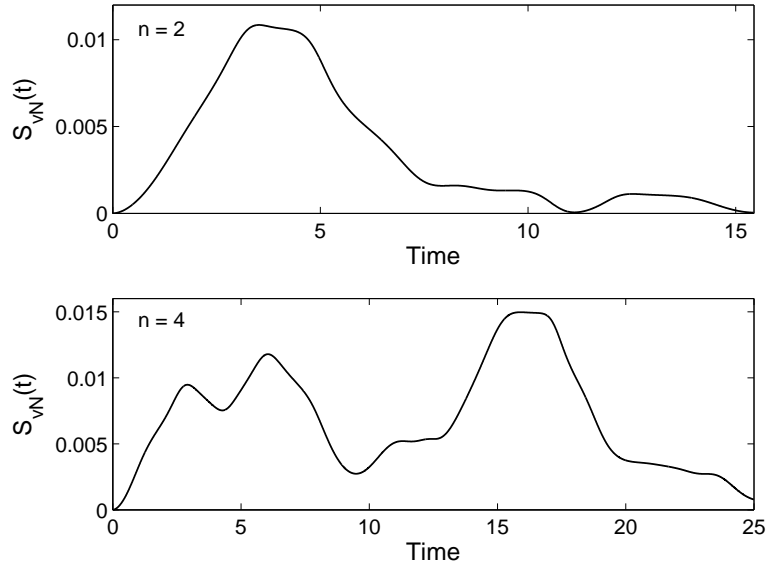}}
\caption{The von Neumann entropy $S_{\mathrm{vN}}(t)$ versus time, for
  the optimally controlled Hadamard gates coupled to two- and
  four-particle environments ($n = 2, 4$, $\gamma = 0.02$, and $\gamma'
  = 0.0175$). The initial state for the entropy computation is
  $|\Psi(0)\rangle$.}
\label{fig:fit-deco}
\end{figure}

Control mechanisms can be better understood by examining the decoherence
dynamics of the qubit. Figure~\ref{fig:fit-deco} shows the time behavior
of the von Neumann entropy of the qubit for the optimally controlled
evolution, with $\gamma = 0.02$, $\gamma' = 0.0175$, and $n = 2, 4$. The
difference between the entropy values for $\gamma' \neq 0$ and $\gamma'
= 0$ is extremely small, implying that the same control mechanism works
in both cases. We observe that the optimal control dramatically enhances
coherence of the qubit system in comparison to the uncontrolled
dynamics. Decoherence is suppressed by the control at all times, but
especially at the end of the transformation. For example, for the
Hadamard gate with $\gamma = 0.02$, $S_{\mathrm{vN}}(t_{\mathrm{f}}) <
10^{-7}$ in the presence of a one-particle environment, which means that
at $t = t_{\mathrm{f}}$ the qubit and environment are almost
uncoupled. While the value of $S_{\mathrm{vN}}(t_{\mathrm{f}})$
increases with the environment size (see table~\ref{tab:results}), the
optimal control is still able to achieve a significant degree of
decoupling at the final time.  Inspecting eigenvalues of the controlled
Hamiltonian, we find that the intense control field creates significant
dynamic Stark shifts of the energy levels. This effect is mainly
responsible for reducing the qubit-environment interaction during the
control pulse. However, achieving extremely low final-time entropies and
correspondingly high gate fidelities requires the employment of an
induced coherence revival. In the uncontrolled system, revivals occur at
times much longer than $t_{\mathrm{f}}$, so that the strong coherence
revival observed at $t = t_{\mathrm{f}}$ is induced exclusively by the
control field. However, as the complexity of the composite system
increases, it becomes more difficult to induce an almost perfect
revival; therefore, the gate fidelity and final-time coherence decrease
as $n$ increases.  Similar results are found for a two-qubit
controlled-NOT gate optimized in the presence of a one-particle
environment, in which case we observe \cite{Grace07} a longer pulse
duration ($t_{\mathrm{f}} = 121.1$) and smaller fidelity ($F \approx
0.9798$) than for one-qubit gates.

\section{Robustness of optimally controlled gates to coupling-strength
  variations}

In realistic quantum systems, the strength of coupling between particles
not always can be accurately measured and is also subject to
fluctuations caused by noise or variations due to imperfect
manufacturing or preparation. Therefore, it is important to explore the
effect of general symmetry-breaking variations in the coupling
parameters $\gamma_{i j}$ on operations of optimally controlled quantum
gates. Given the one-qubit Hadamard gate as the target transformation
and a fixed number $n$ of environmental particles ($n = 2, 4$), we find
the optimal control field for a specified set of coupling parameters
$\gamma_{ij}$ of Eq.~(\ref{gamma}), with $\gamma = 0.02$ and $\gamma' =
c \gamma$ ($c = 0, \frac{1}{2}, \frac{7}{8}$). Then we apply this
control field to an ensemble of systems with normally distributed
variations in the coupling parameters $\gamma_{ij}$ and analyze how the
uncertainty in the coupling strengths affects the gate fidelity and
final-time entropy. Although the dependence of $F$ and
$S_{\mathrm{vN}}(t_{\mathrm{f}})$ on the coupling parameters is
non-linear (which implies that the corresponding distributions of $F$
and $S_{\mathrm{vN}}(t_{\mathrm{f}})$ will not be normal), our
statistical analysis employs only mean values and standard deviations,
given by $\overline{F} = L^{-1} \sum_{p = 1}^{L} F_p$ and $\sigma_F = [
L^{-1} \sum_{p = 1}^{L} ( F_p - \overline{F} )^2 ]^{1/2}$, respectively,
for the gate fidelity $F$, and similarly for the final-time entropy
$S_{\mathrm{vN}}(t_{\mathrm{f}})$. The summation is over all elements of
the ensemble (ensemble sizes $L$ of the order of $10^5$ are used in the
calculations).

\begin{figure}[!ht]
\epsfxsize=0.62\textwidth \centerline{\epsfbox{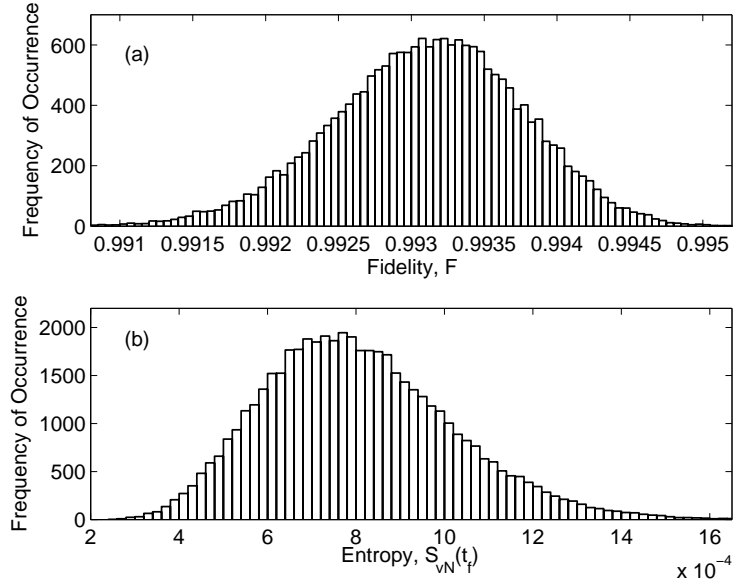}}
\caption{Frequency histograms for (a) the gate fidelity distribution
  and (b) the final-time entropy distribution. These distributions are
  obtained when the control field optimized for the Hadamard gate with
  $n = 4$, $\gamma = 0.02$, and $\gamma' = (7/8)\gamma = 0.0175$ is
  applied to an ensemble of systems with normally distributed variations
  in the coupling parameters $\gamma_{ij}$, as explained in the text.
  Table~\ref{tab:robust} reports statistical data for these
  distributions. Note that the sub-plots have different scales of the
  axes.}
\label{fig:RFO}
\end{figure}

For each element of the statistical ensemble, the value of each
qubit-environment coupling parameter $\gamma_{0j}$ ($j = 1, \ldots, n$)
is randomly selected from the normal distribution with a mean
$\overline{\gamma} = 0.02$ and a standard deviation $\sigma_{\gamma} =
\overline{\gamma}/8 = 0.0025$. Analogously, for non-zero
inter-environment coupling,\footnote{In the case of zero
inter-environment coupling, $c = 0$, all zero values of $\gamma_{ij}$
are left unchanged.} the value of each coupling parameter $\gamma_{ij}$
($i = 1, \ldots, n-1$, $j > i$) is also randomly selected from the
normal distribution with a mean $\overline{\gamma'} = c \gamma$ and a
standard deviation $\sigma_{\gamma'} = \overline{\gamma'}/8 = c
\overline{\gamma}/8$. The statistical analysis of the corresponding
fidelity and final-time entropy distributions is reported in
table~\ref{tab:robust} (for $n = 2,4$), and frequency histograms of
these distributions are shown in figure~\ref{fig:RFO} (for $n =
4$). These results demonstrate a high degree of robustness of the
optimally controlled gates to relatively large variations in the
qubit-environment and inter-environment coupling strengths. For given
values of $n$, $\overline{\gamma}$, and $\overline{\gamma'}$, on average
there is just a minuscule decrease in the fidelity and final-time
coherence due to the coupling strength variations, and the relative
width of the fidelity distribution, $\sigma_F / \overline{F}$, is
smaller than $\sigma_{\gamma} / \overline{\gamma}$ by several orders of
magnitude. Based on the data in table~\ref{tab:robust}, we also observe
that the standard deviations $\sigma_F$ and $\sigma_{S_{\mathrm{vN}}}$
increase with the number of the environmental
particles. Table~\ref{tab:robust} also helps us to see that
inter-environment couplings have very little effect on the gate
performance, although their influence slightly increases with
$\overline{\gamma'}$.

\begin{table}
\tbl{Fidelity and entropy data for the one-qubit Hadamard gate, obtained
  when the control field optimized for a specified set of the system
  parameters is applied to an ensemble of systems with normally
  distributed  variations in the coupling parameters
  $\gamma_{ij}$. Columns of $F$ and $S_{\mathrm{vN}}(t_{\mathrm{f}})$
  contain fidelity and final-time entropy values, respectively, obtained
  for the specified coupling strengths: $\gamma = 0.02$ and $\gamma' = c
  \gamma$ ($c = 0, 1/2, 7/8$). Columns of $\overline{F}$ and
  $\overline{S_{\mathrm{vN}}}$ contain mean values of the fidelity and
  final-time entropy, respectively, over the ensemble, while $\sigma_F$
  and $\sigma_{S_{\mathrm{vN}}}$ are the respective standard
  deviations.}
{\begin{tabular}{@{}ccccccc} 
\toprule 
${\bm{n = 2}}$ & & & & & & \\ 
$\gamma'$ & $F$ & $\overline{F}$ & $\sigma_F$ &
  $S_{\mathrm{vN}}(t_{\mathrm{f}})$ & $\overline{S_{\mathrm{vN}}}$ &
  $\sigma_{S_{\mathrm{vN}}}$ \\ 
\colrule 
$\gamma' = 0$ & 0.9975 & 0.9975 & $2.6 \times 10^{-4}$ & $4.4 \times
  10^{-5}$ & $4.6 \times 10^{-5}$ & $1.5 \times 10^{-5}$ \\ 
\colrule 
$\overline{\gamma'} = (1/2)\overline{\gamma}$ & 0.9975 & 0.9975 & $2.6
  \times 10^{-4}$ & $4.5 \times 10^{-5}$ & $4.6 \times 10^{-5}$ & $1.5
  \times 10^{-5}$ \\ 
$\overline{\gamma'} = (7/8)\overline{\gamma}$ & 0.9975 & 0.9975 & $2.6
  \times 10^{-4}$ & $4.5 \times 10^{-5}$ & $4.7 \times 10^{-5}$ & $1.5
  \times 10^{-5}$ \\  \toprule ${\bm{n = 4}}$ & & & & & & \\ 
$\gamma'$ & $F$ & $\overline{F}$ & $\sigma_F$ &
  $S_{\mathrm{vN}}(t_{\mathrm{f}})$ & $\overline{S_{\mathrm{vN}}}$ &
  $\sigma_{S_{\mathrm{vN}}}$ \\
\colrule 
$\gamma' = 0$ & 0.9935 & 0.9934 & $6.1 \times 10^{-4}$ & $4.7 \times
  10^{-4}$ & $4.8 \times 10^{-4}$ & $8.1 \times 10^{-5}$ \\
\colrule 
$\overline{\gamma'} = (1/2)\overline{\gamma}$ & 0.9934 & 0.9933 & $6.3
  \times 10^{-4}$ & $6.4 \times 10^{-4}$ & $6.5 \times 10^{-4}$ & $1.4
  \times 10^{-4}$ \\  
$\overline{\gamma'} = (7/8)\overline{\gamma}$ & 0.9933 & 0.9931 & $6.5
  \times 10^{-4}$ & $7.8 \times 10^{-4}$ & $8.1 \times 10^{-4}$ & $2.1
  \times 10^{-4}$ \\
  \botrule
\end{tabular}}
\label{tab:robust} 
\end{table}

\section{Conclusions}

This work demonstrates the importance of the optimal control theory for
designing quantum gates, especially in the presence of
environmentally-induced decoherence. The model studied here represents a
realistic system of interacting qubits with uncertain coupling strengths
and is relevant for various physical implementations of QC. Very precise
optimal solutions obtained in the presence of unwanted couplings reveal
control mechanisms which employ fast and intense time-dependent fields
to effectively suppress decoherence via dynamic Stark shifting and
restore coherence via an induced revival. In addition, these optimal
solutions exhibit a significant degree of inherent robustness to random
variations in the coupling strengths. It is also found that optimally
controlled gate operations are practically unaffected by interactions
between the environmental particles. These results further support the
use of laboratory closed-loop optimal controls in QC applications.

\section*{Acknowledgments}

This work was supported by the ARO-QA, DOE, and NSF. D.~A.~L. was
supported by ARO-QA Grant No.~W911NF-05-1-0440 and NSF Grant
No.~CCF-0523675. I.~A.~W. acknowledges support by the UK QIP IRC funded
by EPSRC, and the EC under the Integrated Project QAP funded by the IST
directorate as Contract No.~015848.



\end{document}